\documentclass[11pt,preprint]{aastex}

\usepackage{epsfig}
\usepackage{amsmath}
\usepackage{natbib}
\usepackage{rotating}
\usepackage{lscape,longtable}

\pdfoutput=1

\begin{document}

\title{Near-Infrared K and L$'$ Flux Ratios in Six Lensed Quasars}

\author{Ross Fadely\altaffilmark{1,2} \& Charles R. Keeton\altaffilmark{1}}
\altaffiltext{1}{Department of Physics and Astronomy, Rutgers, the State University of New Jersey, 136 Frelinghuysen Road, Piscataway, NJ 08854; keeton@physics.rutgers.edu}
\altaffiltext{2}{(Current) Department of Astronomy, Haverford College, 370 Lancaster Ave., Haverford, PA 19041;
rfadely@haverford.edu}

%%%%%%%%%%%%%%%%%%%%%%%%%%%%%%%%%%%%%%%%%%%%%%%%%%%%%%%%%%%%%%%%%

\begin{abstract}

%%%%%%%%%%%%%%%%%%%%%%%%%%%%%%%%%%%%%%%%%%%%%%%%%%%%%%%%%%%%%%%%%

We examine the wavelength dependence of flux ratios for six gravitationally lensed quasars using $K$ and $L'$ images obtained at the Gemini North 8m telescope.  We select lenses with source redshifts $z_s < 2.8$ so that $K$-band images probe rest-frame optical emission from accretion disks, while $L'$-band images probe rest-frame near-infrared flux emitted (in part) from the more extended surrounding torus.  Since the observations correspond to different source sizes, the $K$ and $L'$ flux ratios are sensitive to structure on different scales and may be useful for studying small-structure in the lens galaxies.  Four of the six lenses show differences between $K$ and $L'$ flux ratios.  In HE 0435$-$1223, SDSS 0246$-$0825, and HE 2149$-$2745 the differences may be attributable to known microlensing and/or intrinsic variability.  In SDSS 0806+2006 the wavelength dependence is not easily attributed to known variations, and may indicate the presence of substructure.  By contrast, in Q0142$-$100 and SBS 0909+523 the $K$ and $L'$ flux ratios are consistent within the uncertainties.  We discuss the utility of the current data for studying chromatic effects related to microlensing, dust extinction, and dark matter substructure.

\end{abstract}

%%%%%%%%%%%%%%%%%%%%%%%%%%%%%%%%%%%%%%%%%%%%%%%%%%%%%%%%%%%%%%%%%

\section{Introduction}
\label{sec:irintro}

%%%%%%%%%%%%%%%%%%%%%%%%%%%%%%%%%%%%%%%%%%%%%%%%%%%%%%%%%%%%%%%%%

While the cold dark matter (CDM) paradigm for structure formation successfully describes cosmological observations on large (CMB and cluster) scales, there is notable disagreement with small-scale observations.  Among other issues, $N$-body simulations (e.g., Via Lactea -- Diemand et al. 2008; Aquarius -- Springel et al. 2008) predict the existence of numerous CDM subhalos, with masses $M \sim 10^4$--$10^9~M_\odot$, embedded in galaxy-scale dark matter halos.  This has proved troubling observationally, because there are many fewer dwarf galaxies in our own Milky Way than predicted by CDM.  Since the discrepancy may be due to baryon stripping from subhalos \citep[e.g.,][]{2008ApJ...679.1260M, 2010MNRAS.402.1995M}, we need ways to probe dark matter substructure directly, regardless of the presence of baryonic material.  

Gravitational lensing provides a unique way to detect CDM substructure in distant galaxies \citep[e.g.,][]{2001ApJ...563....9M, 2002ApJ...565...17C,2002ApJ...572...25D}.	Stars and CDM substructure perturb the lens potential on micro- to milli-arcsecond scales, which can have dramatic effects on the properties of lensed images.   Most notably, lensing from stars and dark matter substructure can alter the flux ratios from those of smooth mass distributions.  As shown by \citet{2006MNRAS.365.1243D}, lens flux ratios depend on the size of the source compared to the size of the perturber.  When the source is very small, it is effectively a point source for the lens substructure, so it feels an additional magnification boost ($B_{\rm sub} \equiv \mu_{\rm sub}/\mu_{\rm smooth}$) over and above the macroscopic lens properties.  As the source increases in size, $B_{\rm sub}$ may increase or decrease depending on the source's location relative to the substructure and the parity of the ``macro'' magnification.  As the source grows still larger, it ceases to be affected by the substructure and $B_{\rm sub} \rightarrow 1$. This phenomenon implies that by measuring flux ratios at different wavelengths, corresponding to different source sizes, we may be able to map substructure on a variety of scales \citep[also see][]{2003MNRAS.339..607M}.

Heuristically, a quasar emitting region of size $R_S$ is significantly affected by a subhalo with Einstein radius $R_E$ only if $R_S \lesssim R_E$.  For typical lens and source redshifts (e.g., $z_l=0.5$ and $z_s=2.0$), the Einstein radius of a subhalo of mass $M$ is $R_E\sim 10^{16}~\mbox{cm}~(M/M_\odot)^{1/2}$.  Since the optically emitting regions of quasars have $R_S \sim 10^{15}$--$10^{16}$ cm \citep{2000MNRAS.315...62W,2007ApJ...661...19P,2010ApJ...712.1129M}, optical lens flux ratios are sensitive to both microlensing by stars and millilensing by CDM substructure.  By contrast, the more extended infrared emitting regions with $R_S \gtrsim 1$ pc \citep{2005ApJ...627...53C, 2009ApJ...697..610M, 2009ApJ...697.1010A} can only be affected by relatively massive subhalos.  Comparing lens flux ratios at different wavelengths therefore makes it possible to constrain the amount of micro- and milli-lensing present in the system, as well as the sizes of the perturbers.

Previous studies have used mid-IR observations to probe rest-frame IR emission, yielding evidence for subhalos with masses $\gtrsim 10^5 M_\Sun$ in the lenses B1422+231 and MG 0414+0534 \citep{2005ApJ...627...53C, 2009ApJ...697..610M}, constraints on the mass of a luminous companion galaxy in H1413+117 \citep{2009ApJ...699.1578M},\footnote{\citet{2009MNRAS.394..174M} also constrained the mass of a luminous satellite, but using radio observations.} and null detections in several other systems \citep[][]{2005ApJ...627...53C, 2009ApJ...697..610M, 2009ApJ...697.1010A}.  Here we extend the study of wavelength-dependent flux ratios by using $K$ ($2.2\,\mu$m) and $L'$ ($3.8\,\mu$m) images of six lenses obtained with Gemini North during the 2008B semester.  For source redshifts $z_s < 2.8$, the $L'$-band images correspond to rest-frame emission at $>1\,\mu$m where $\sim$20--100\% of the flux is thermal emission from the inner dusty torus \citep{1995MNRAS.272..737R, 2004ApJ...600L..35M, 2008ApJ...685..160N, 2009ApJ...697.1010A}.  By contrast, the $K$-band flux comes mostly from the smaller accretion disk.  Thus, comparing $K$ and $L'$ flux ratios may provide a sufficient source size baseline to identify substructure. 

To be sure, there are phenomena besides CDM substructure that may cause lens flux ratios to vary with wavelength.  Optical observations probe rest-frame UV emission from the accretion disk of the AGN, so they are sensitive to microlensing by stars in the lens galaxy \citep[e.g.,][]{2009ApJ...693..174C, 2010ApJ...712.1129M}.  Optical flux ratios can also be altered by differential extinction from dust in the lens galaxy \citep[e.g.,][]{2006ApJS..166..443E}.  Finally, intrinsic variability in the source coupled with the lens time delay can cause flux ratios to vary with time, and the variability may be chromatic.\footnote{Photometric monitoring can be used to quantify the variations, but has only been done for certain lenses \citep{2006A&A...451..747E,2006ApJ...640...47K,2007A&A...464..845V,2008A&A...488..481V,2008ARep...52..270K,2008NewA...13..182G}.  Four of our targets have been monitored at optical wavelengths: Q0142$-$100, HE 0435$-$1223, SBS 0909+523, and HE 2149$-$2745.}  All three effects should be attenuated at near-IR wavelengths, because the effective size of the accretion disk is larger and some of the flux originates in the larger dusty torus, but they may not be entirely absent.  In particular, the importance of microlensing and intrinsic variability for $L'$-band observations will depend on the relative strengths of the accretion disk (0.01-- 0.1 pc, $10^{-6}$--$10^{-5}$ arcsec) and dusty torus (0.5--5 pc, $10^{-4}$--$10^{-3}$ arcsec) emission.  Any significant wavelength-dependence in lens flux ratios is interesting, whether related to CDM substructure or not, so at minimum our observations can highlight systems that warrant further study.

Most lensing constraints on CDM substructure have relied on the identification of flux ratio anomalies, which is best done using four-image lenses.  The reason is that identifying such anomalies requires either well-constrained lens models \citep[e.g.,][]{2002ApJ...572...25D, 2002ApJ...567L...5M} or universal magnification relations that apply only to certain four-image configurations \citep{2003ApJ...598..138K, 2005ApJ...635...35K}.  However, the search for wavelength dependence in flux ratios is a purely observational task that may provide model-independent evidence for substructure.  We are therefore able to analyze two-image lenses for the presence of substructure for the first time.  Some care is needed when interpreting the results (see \S \ref{sec:irdiscuss}), but the observations are still valuable.

%%%%%%%%%%%%%%%%%%%%%%%%%%%%%%%%%%%%%%%%%%%%%%%%%%%%%%%%%%%%%%%%%

\section{Observations}
\label{sec:irobs}

%%%%%%%%%%%%%%%%%%%%%%%%%%%%%%%%%%%%%%%%%%%%%%%%%%%%%%%%%%%%%%%%%

To explore the multi-wavelength behavior of flux ratios, we obtained the first $K$ ($2.2\,\mu$m) and $L'$ ($3.8\,\mu$m) images for six gravitational lenses using the Near Infrared Imager (NIRI) located at the Gemini North facility.  Our lens sample was selected from the set of known quasar lenses that can be found in the CASTLES database.\footnote{http://www.cfa.harvard.edu/castles/}  Our targets (Table \ref{tab:obs}) were chosen to have relatively bright ($m_H<18.5$) and well separated ($>1''$) images, and to be accessible during the 2008B semester.  The targets were also selected to have source redshifts $z_s<2.8$ so that $L'$-band observations correspond to rest-frame emission at $>1.0\,\mu$m.  Information about the observations is given in Table \ref{tab:obs}.

\begin{deluxetable}{ccccccc}
\tablecaption{Targets and Observational Information}
\tablewidth{0pt}
\tablehead{
& & & & & \multicolumn{2}{c}{Total Integration (min)} \\
Target & $z_{lens}$ & $z_{source}$ & $\Delta\theta$ ($''$) & Date(s)\tablenotemark{\dag} & $K$ & $L'$
}
\startdata
Q0142$-$100            & 0.491	& 2.719 & 2.24	& 8.21.2008 			& 3 		& 32  \\
SDSS 0246$-$0825 & 0.723	& 1.680 & 1.04	& 8.18.2008, 8.20.2008 	& 4.8 	& 76  \\
HE 0435$-$1223      	& 0.455	& 1.689 & 2.42	& 8.31.2008,12.22.2008  	& 4.8 	& 64 \\                           
SDSS 0806+2006 	& 0.573	& 1.540 & 1.40	&12.22.2008 			& 2.4 	& 20  \\
SBS 0909+523     	& 0.830	& 1.376 & 1.17	&12.22.2008 			& 1.6 	& 2 \\
HE 2149$-$2745      	& 0.603	& 2.033 & 1.70	&12.22.2008 			& 3 		& 32
\enddata
\tablenotetext{\dag}{Hawaii Standard Time}
\label{tab:obs}
\end{deluxetable}

All of our targets have previous ground- and space-based imaging.  Results from single-epoch \textit{Hubble Space Telescope} (HST) $V$, $I$, and $H$ band photometry have been used in the past to derive lens models \citep[e.g.,][]{2000ApJ...536..584L}.  Among the various ground-based studies, optical monitoring programs are particularly useful to us because they help quantify the intrinsic variability in the source and the amplitude of variability induced by microlensing \citep{2006ApJ...640...47K, 2008ARep...52..270K, 2008NewA...13..182G}.

At wavelengths $\gtrsim 2\,\mu$m, thermal emission from Earth's atmosphere becomes an increasingly significant component of the flux recorded at the telescope.  In order to remove the (often dominant) emission from the foreground sky, observations in the infrared are nodded between positions on and off source.  Since our targets are much smaller than the field of view (image separation $\Delta\theta < 3''$, compared with a $22''$ field of view in $L'$), we decided to ``nod on chip'' by using a small $2\times2$, $6''$ dither pattern.  For each dither position, three other positions are available to construct sky model images.  This strategy effectively doubles our observing efficiency by eliminating the need for off-source positions.  In addition to providing sky subtraction images, dithering reduces the sensitivity of the observations to hot or bad pixels in the NIRI array.  

For $K$-band observations, single exposures of 20--40 s were selected to ensure that the bright quasar images (extrapolated from 1.6 to $2.2\,\mu$m assuming $f_\nu\propto\nu^{-0.5}$; \citealt{2005ApJ...631..163S}) would not fill more than 50\% of the detector well.  In the $L'$ band, the sky brightness restricts exposure times to 1 s in order to prevent saturation.  To accumulate the necessary integration time, the short $L'$ exposures were co-added in 30 s blocks at each dither position.

\begin{figure*}[!ht]
%\centering
$K$ Band\hspace{5.5cm}$L'$ Band
\vspace{0.2cm}

\fbox{\includegraphics[clip=true, trim=2.5cm 8.cm 3.cm 6.75cm, width=6.5cm]{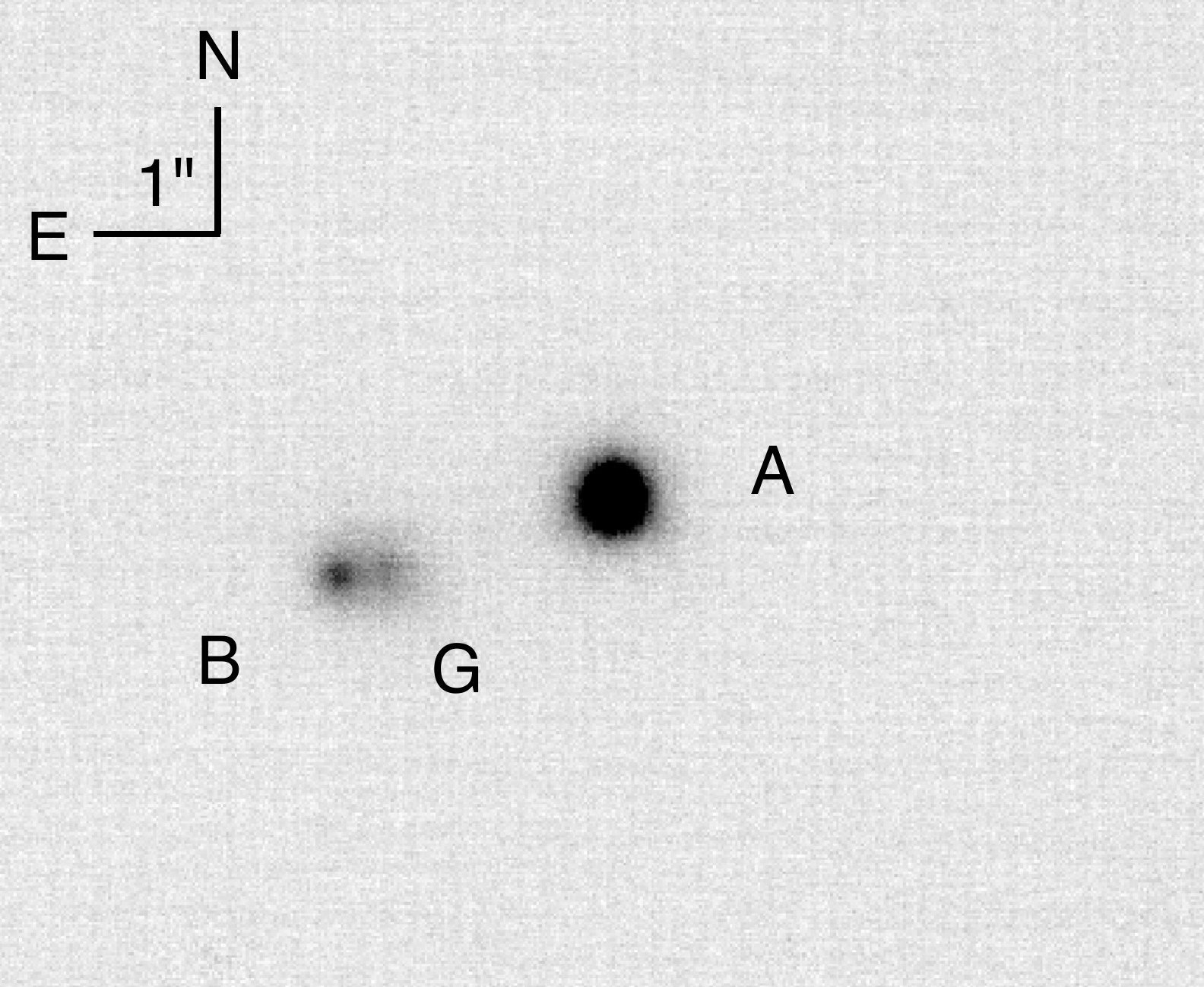}}
\fbox{\includegraphics[clip=true, trim=2.5cm 8.cm 3.cm 6.75cm, width=6.5cm]{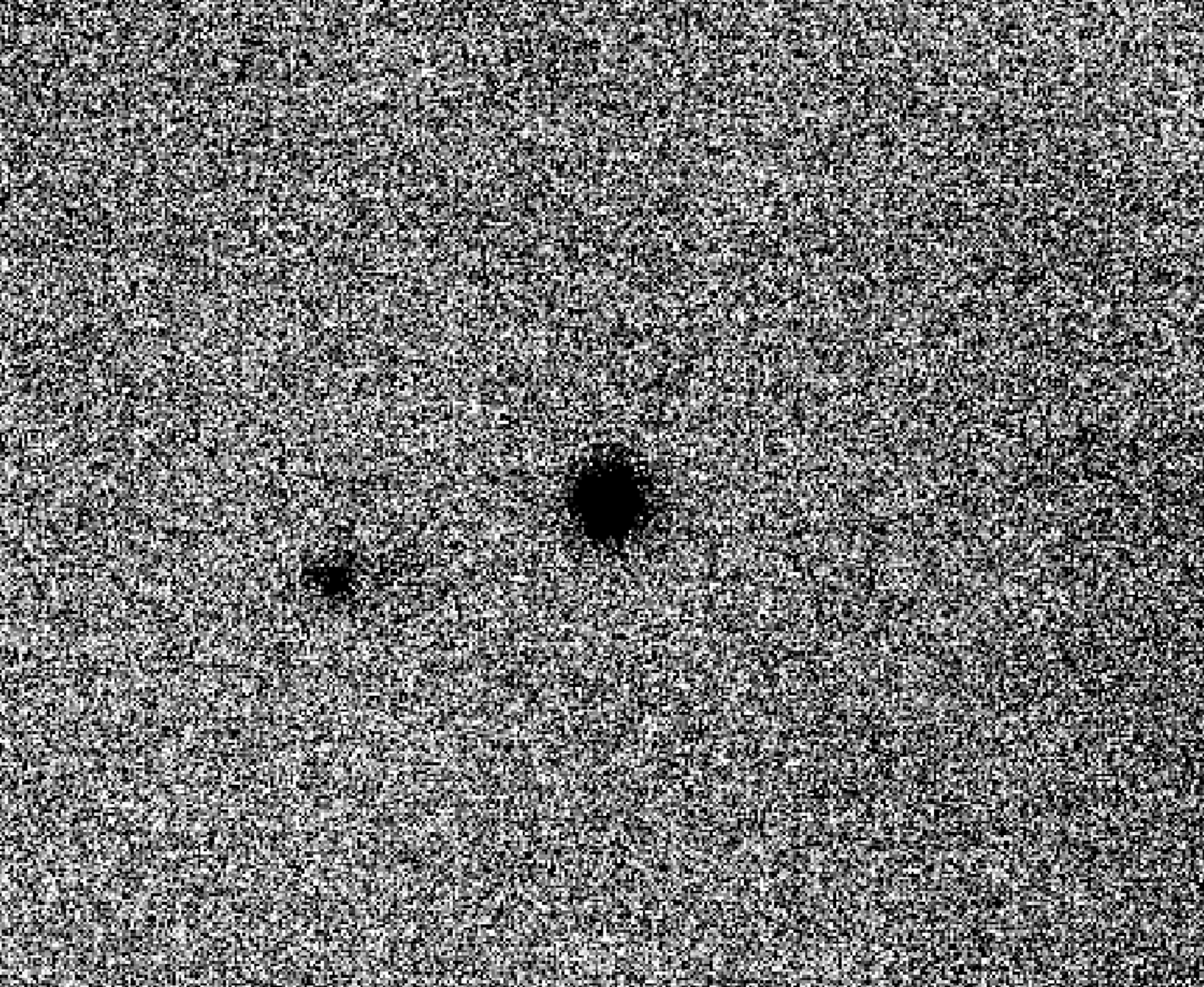}}\\

\fbox{\includegraphics[clip=true, trim=2.5cm 8.cm 3.cm 6.75cm, width=6.5cm]{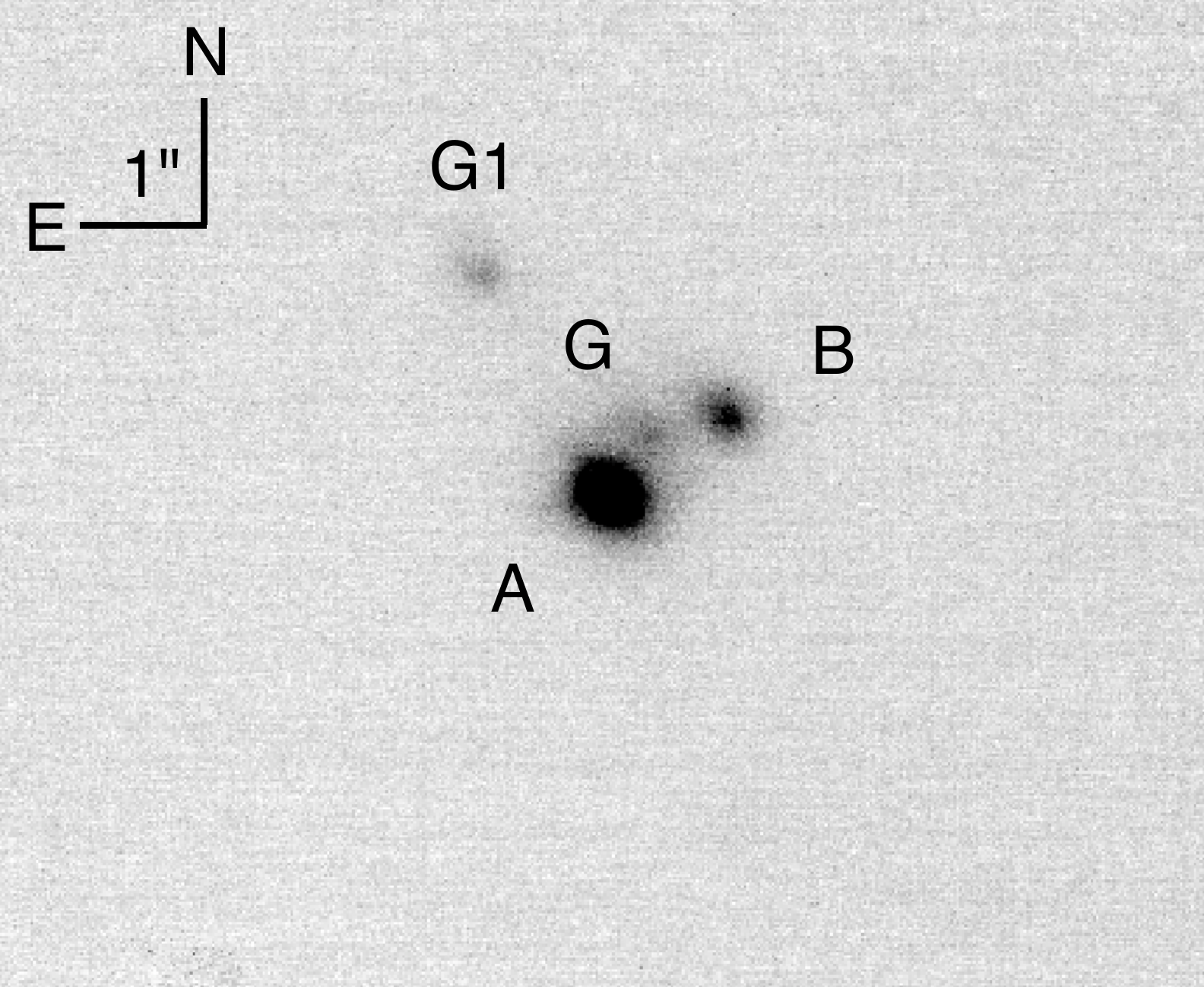}}
\fbox{\includegraphics[clip=true, trim=2.5cm 8.cm 3.cm 6.75cm, width=6.5cm]{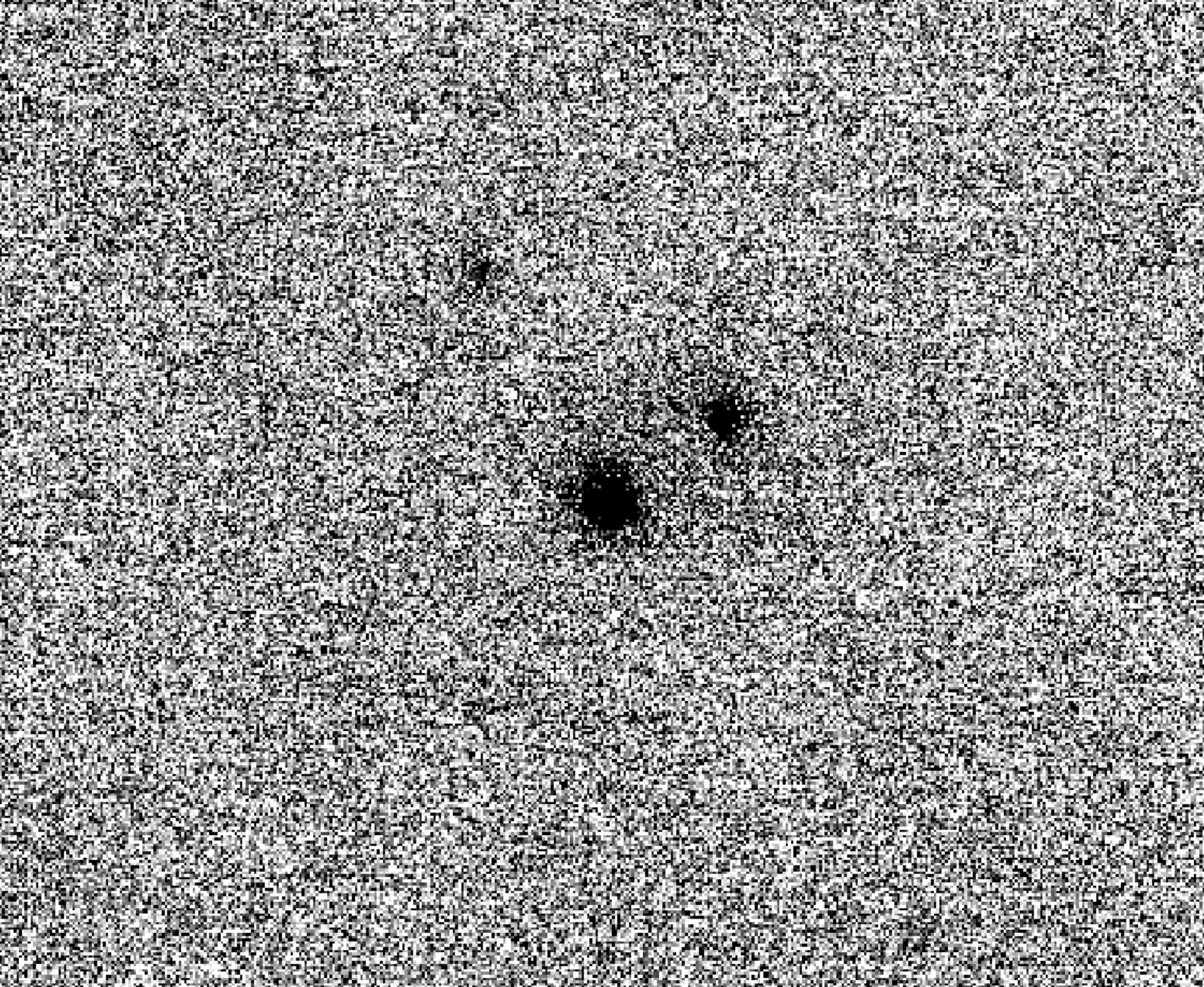}}\\

\fbox{\includegraphics[clip=true, trim=2.5cm 8.cm 3cm 6.75cm, width=6.5cm]{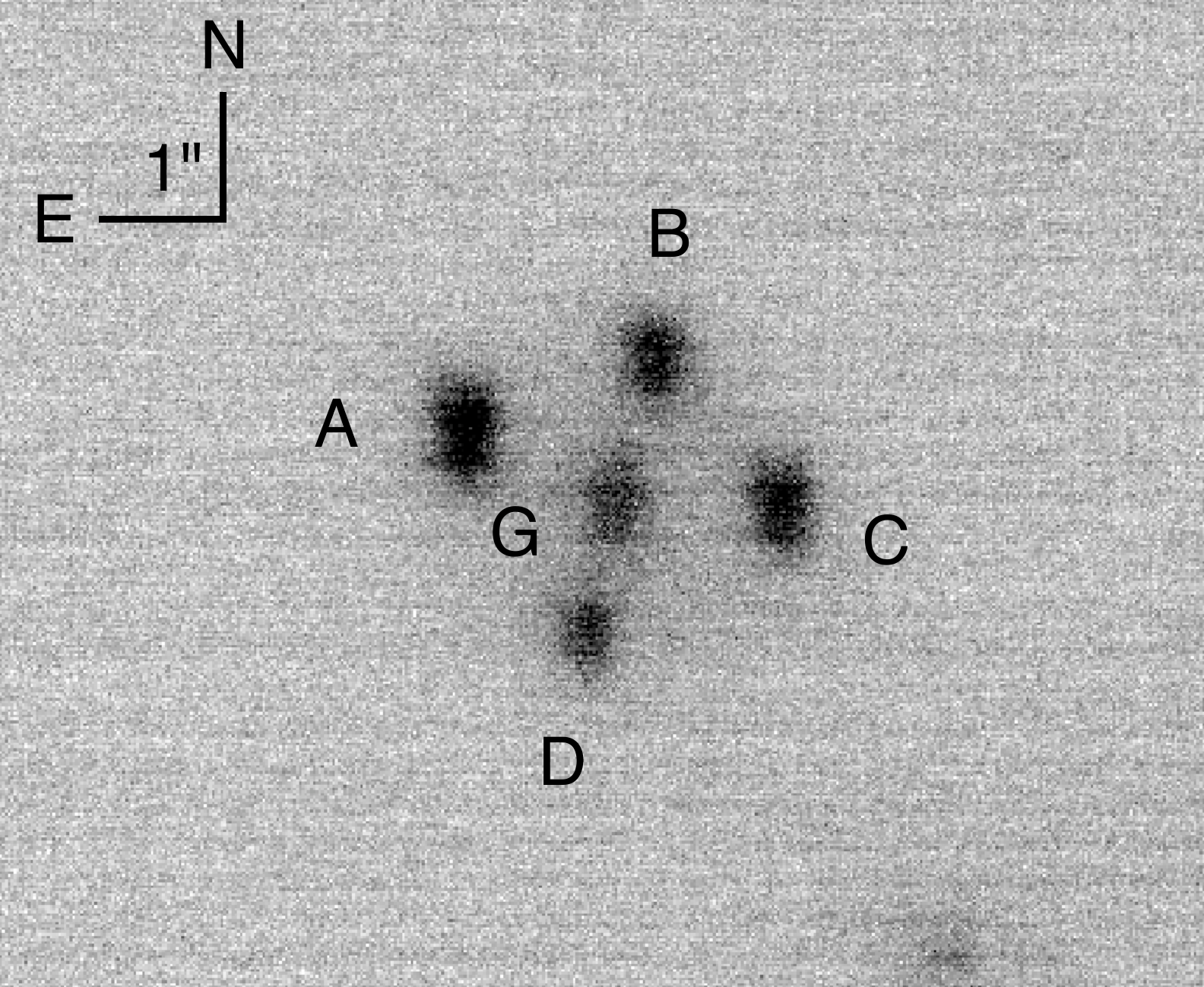}}
\fbox{\includegraphics[clip=true, trim=2.5cm 8.cm 3cm 6.75cm, width=6.5cm]{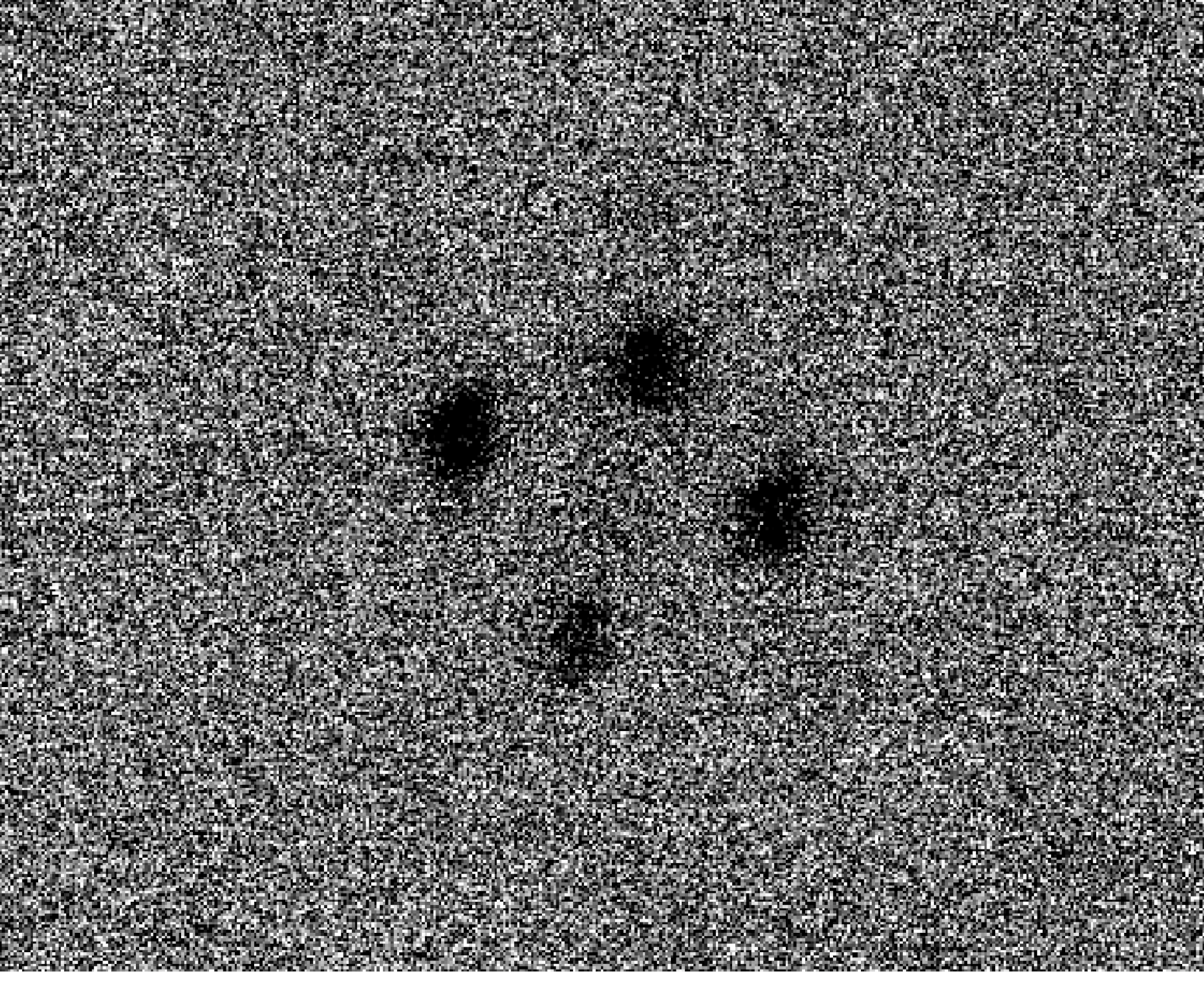}}\\
\caption{$K$ (left) and $L'$ (right) images of Q0142$-$100, SDSS 0246$-$0825, and HE 0435$-$1223 (top to bottom).  Lensed images are labeled as A--D and lens galaxies are labeled as G or G1, following nomenclature in the literature.  Lens galaxies are clearly visible in the $K$ band images.  With the exception of G1 in SDSS 0246$-$0825, lens galaxies are not seen the $L'$ images.}
\label{fig:data1}

\end{figure*}

\begin{figure*}[!ht]
%\centering
$K$ Band\hspace{5.5cm}$L'$ Band
\vspace{0.2cm}

\fbox{\includegraphics[clip=true, trim=2.5cm 8.cm 3cm 6.75cm, width=6.5cm]{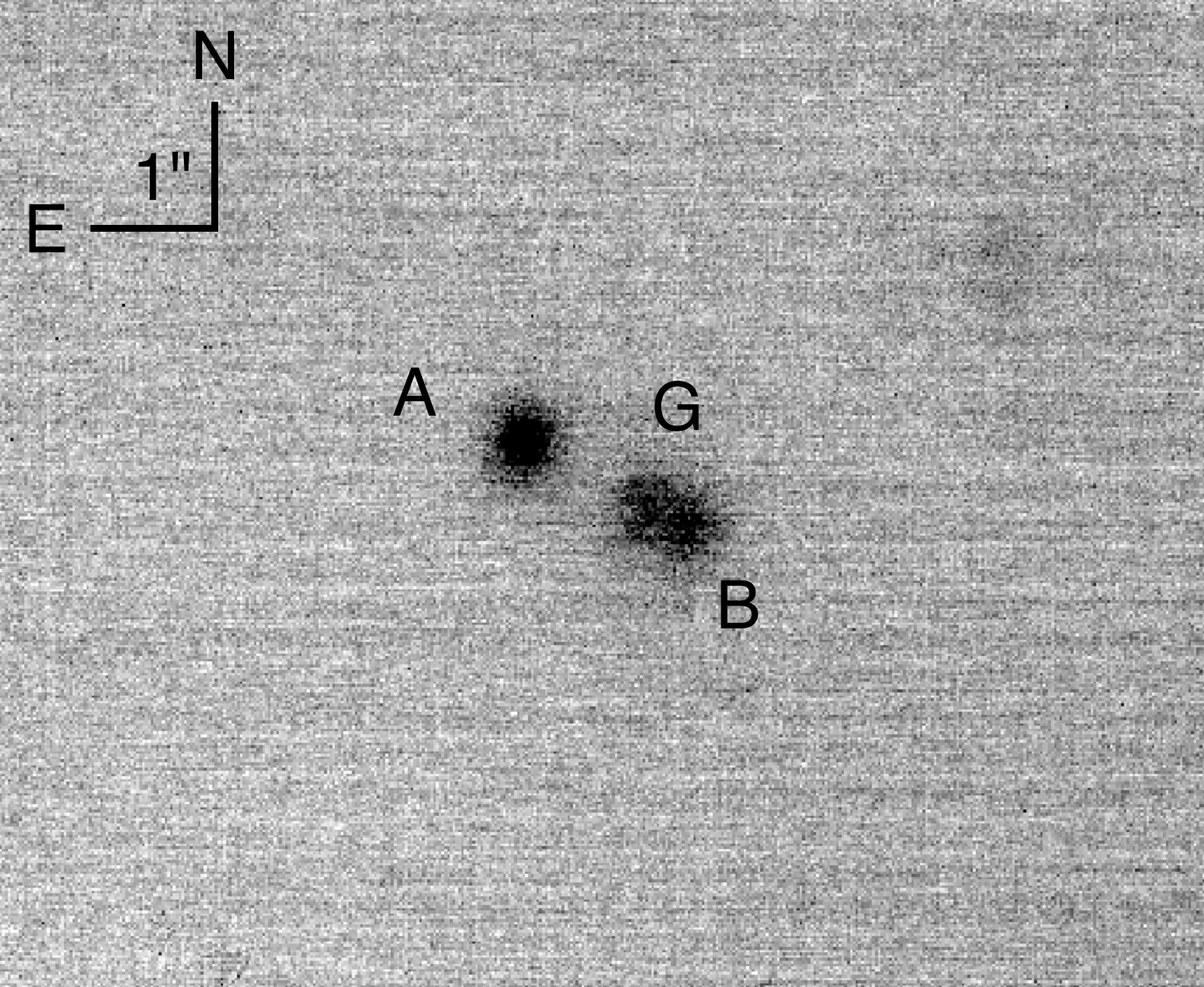}}
\fbox{\includegraphics[clip=true, trim=2.5cm 8.cm 3cm 6.75cm, width=6.5cm]{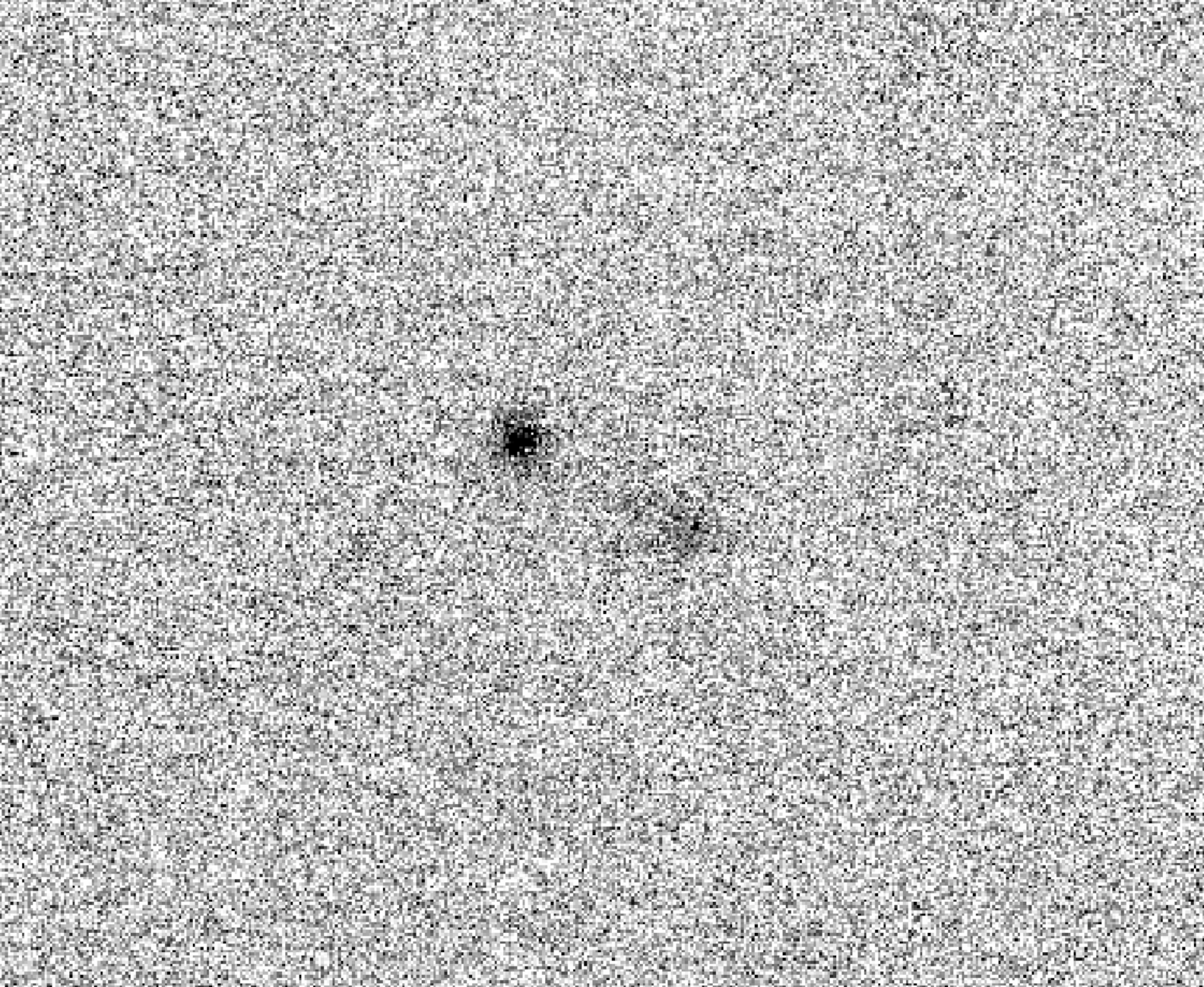}}\\

\fbox{\includegraphics[clip=true, trim=2.5cm 8.cm 3cm 6.75cm, width=6.5cm]{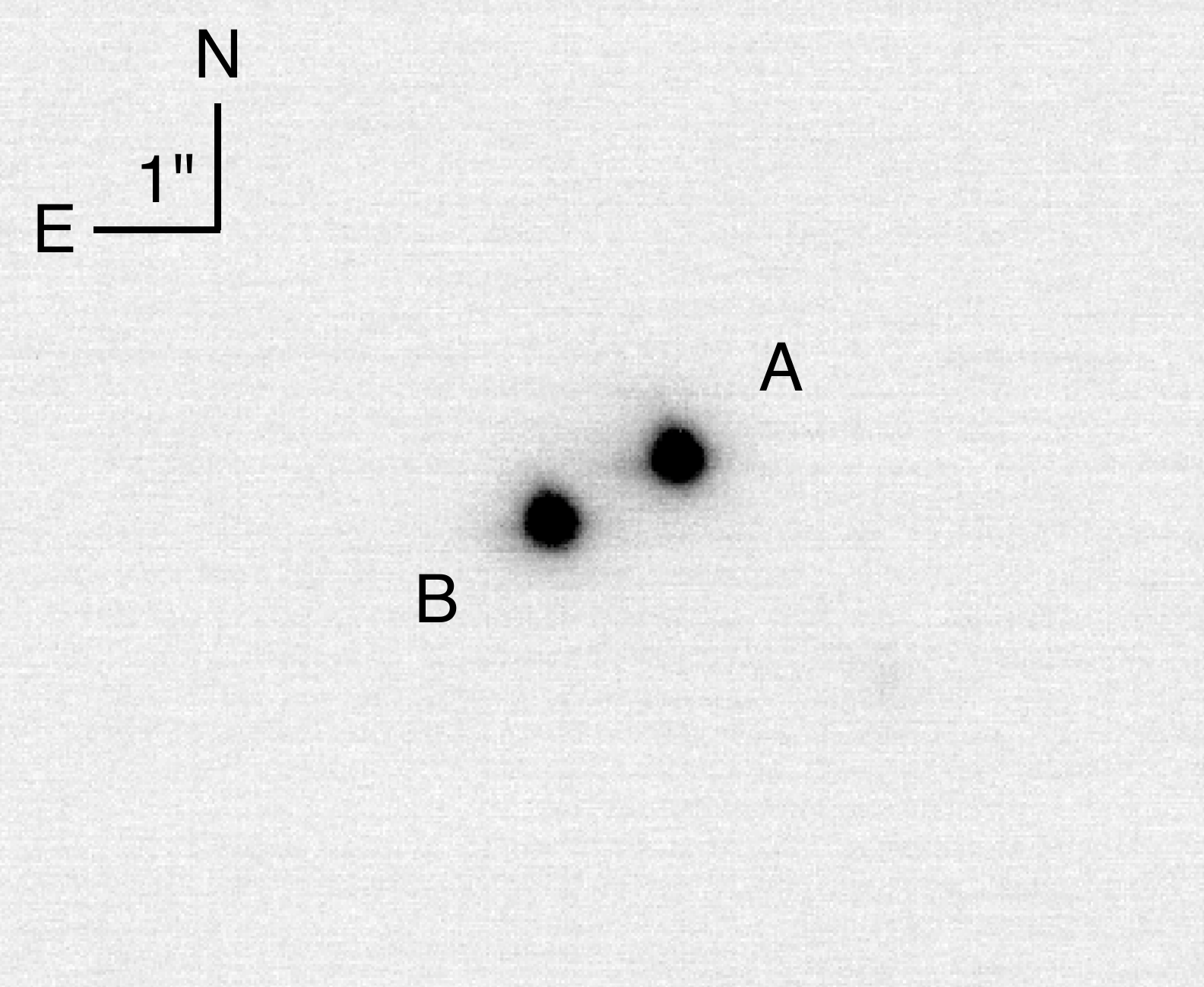}}
\fbox{\includegraphics[clip=true, trim=2.5cm 8.cm 3cm 6.75cm, width=6.5cm]{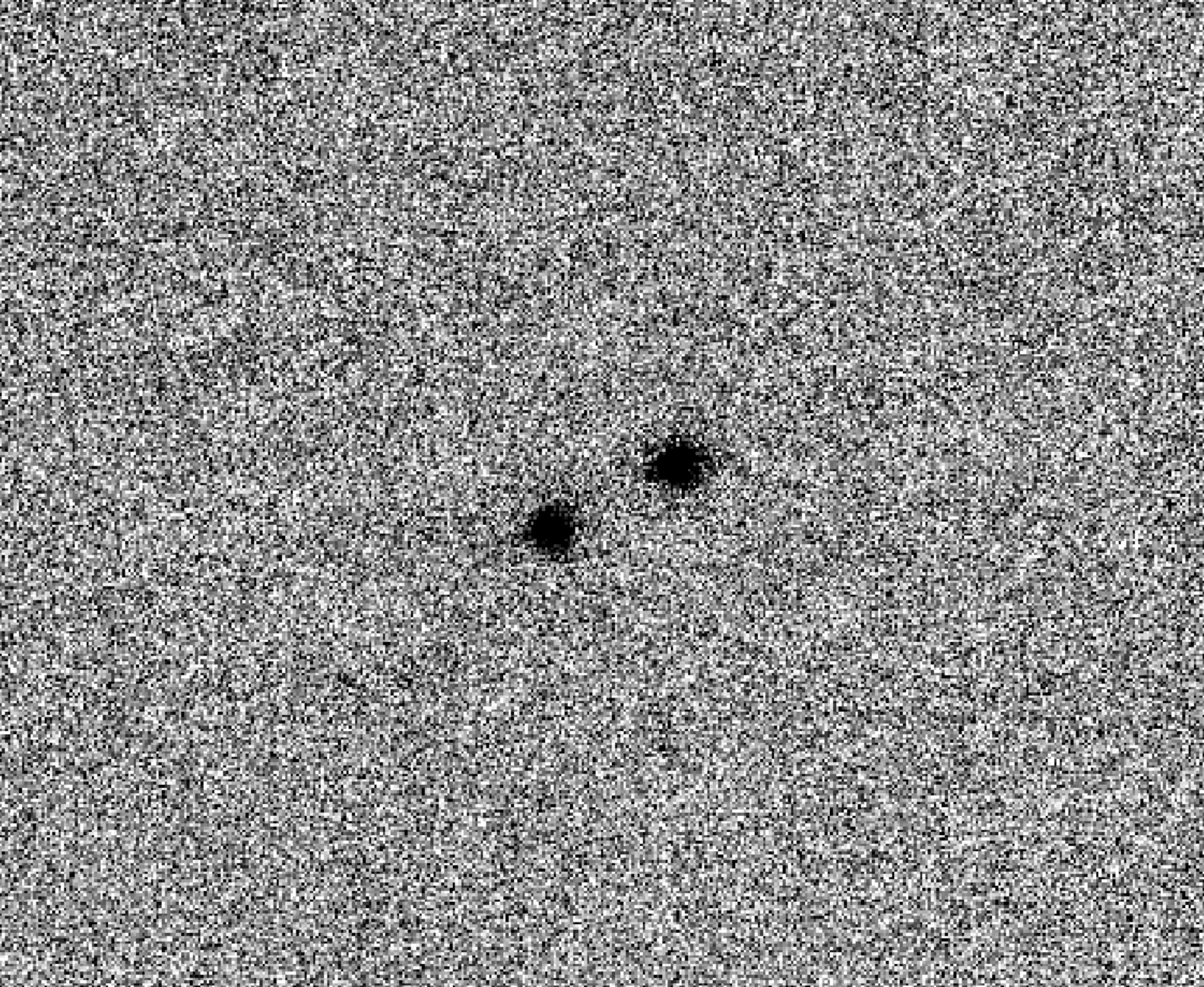}}\\

\fbox{\includegraphics[clip=true, trim=2.5cm 8.cm 3cm 6.75cm, width=6.5cm]{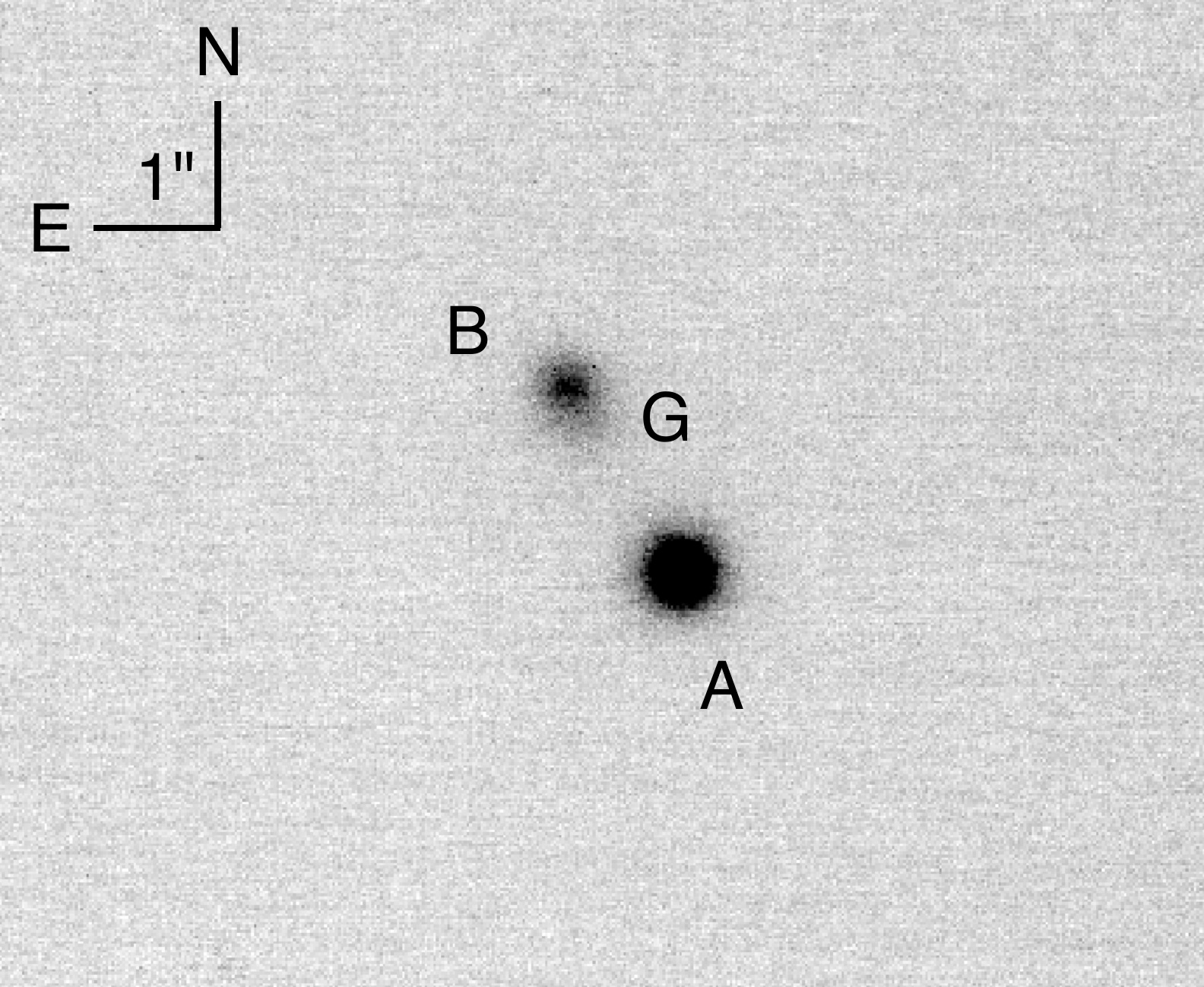}}
\fbox{\includegraphics[clip=true, trim=2.5cm 8.cm 3cm 6.75cm, width=6.5cm]{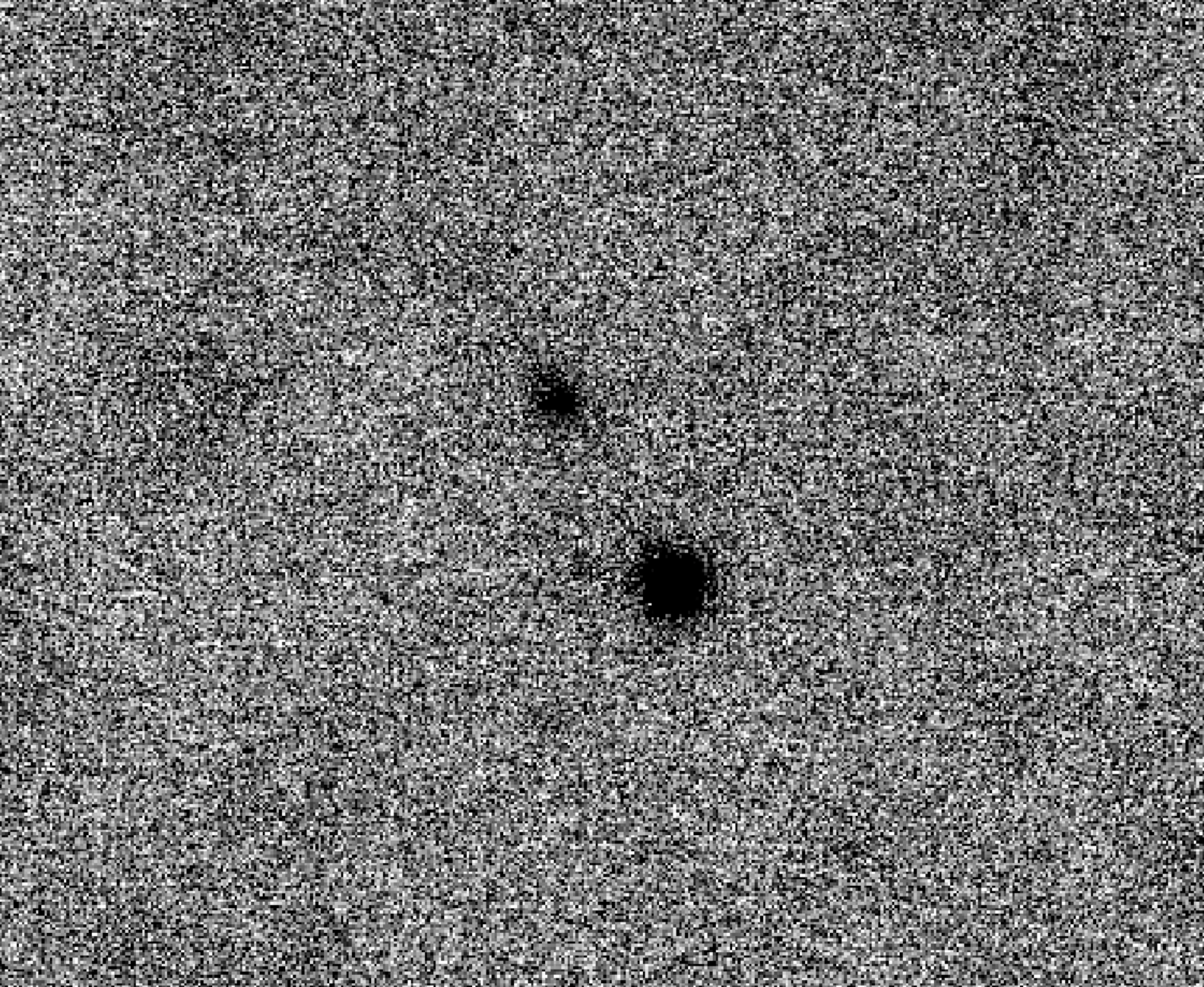}}\\
\caption{Similar to Figure \ref{fig:data1}, but for the lenses SDSS 0806+2006, SBS 0909+523, and HE 2149$-$2745 (top to bottom).}
\label{fig:data2}
\end{figure*}

Data reduction was conducted with standard IRAF routines provided by Gemini Observatory.  The reduction proceeded as follows.  First bad pixel masks and sigma clipping algorithms were applied to account for defective and hot pixels in the detector.  The images were then flat fielded.  For $K$-band observations, sky brightness is subdominant and separate calibration flats were taken each night.  Since the sky dominates $L'$-band observations, the images themselves are essentially sky flats and are used to construct flat field images.  After flat fielding, the data were sky-subtracted using observations from other positions in the same dither pattern.  $L'$ sky images were limited to dither positions immediately before and after a given observation, since sky brightness can vary on the scale of minutes at $3.8\,\mu$m \citep{glass}.  Once each dither position was flat fielded, sky subtracted, and sigma clipped, the final image was produced by combining the various dither positions.  Figures \ref{fig:data1} and \ref{fig:data2} show the reduced $K$ and $L'$ images for our six targets.


%%%%%%%%%%%%%%%%%%%%%%%%%%%%%%%%%%%%%%%%%%%%%%%%%%%%%%%%%%%%%%%%%

\section{Photometry}
\label{sec:irflux}

%%%%%%%%%%%%%%%%%%%%%%%%%%%%%%%%%%%%%%%%%%%%%%%%%%%%%%%%%%%%%%%%%

We use the \texttt{GALFIT} image modeling package \citep{2002AJ....124..266P} to extract the $K$ and $L'$ fluxes of the lensed quasar images and (when detected) the lensing galaxy.  \texttt{GALFIT} is valuable for general photometry, and specifically for separating the flux of the lensed images from that of the nearby lens galaxy, which is detected in our $K$-band images for all targets except SBS 0909+523.  

\begin{figure*}[!ht]
\centering
\fbox{\includegraphics[clip=true, trim=4cm 9cm 4.5cm 7cm, width=6.cm]{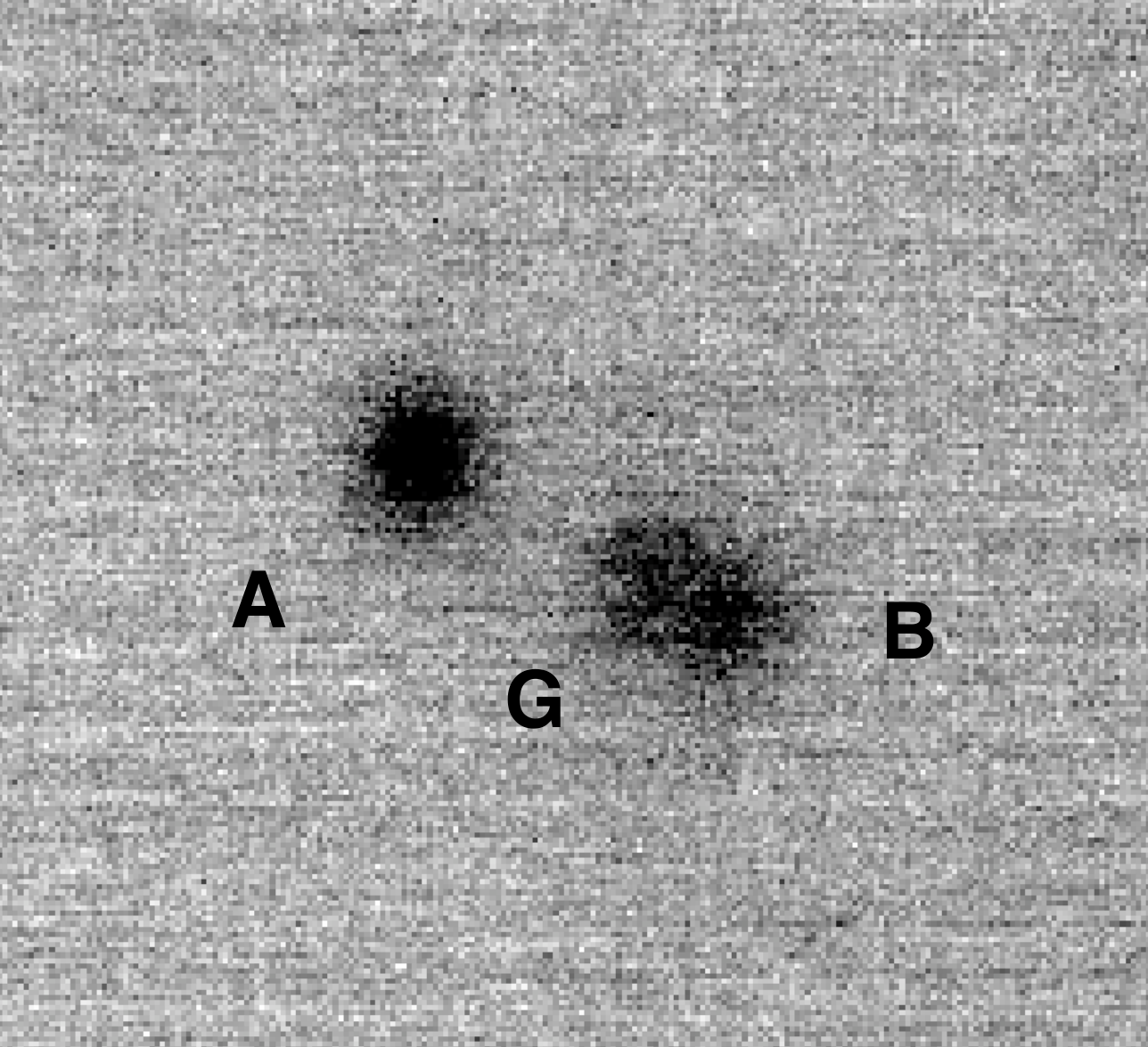}}
\fbox{\includegraphics[clip=true, trim=4cm 9cm 4.5cm 7cm, width=6.cm]{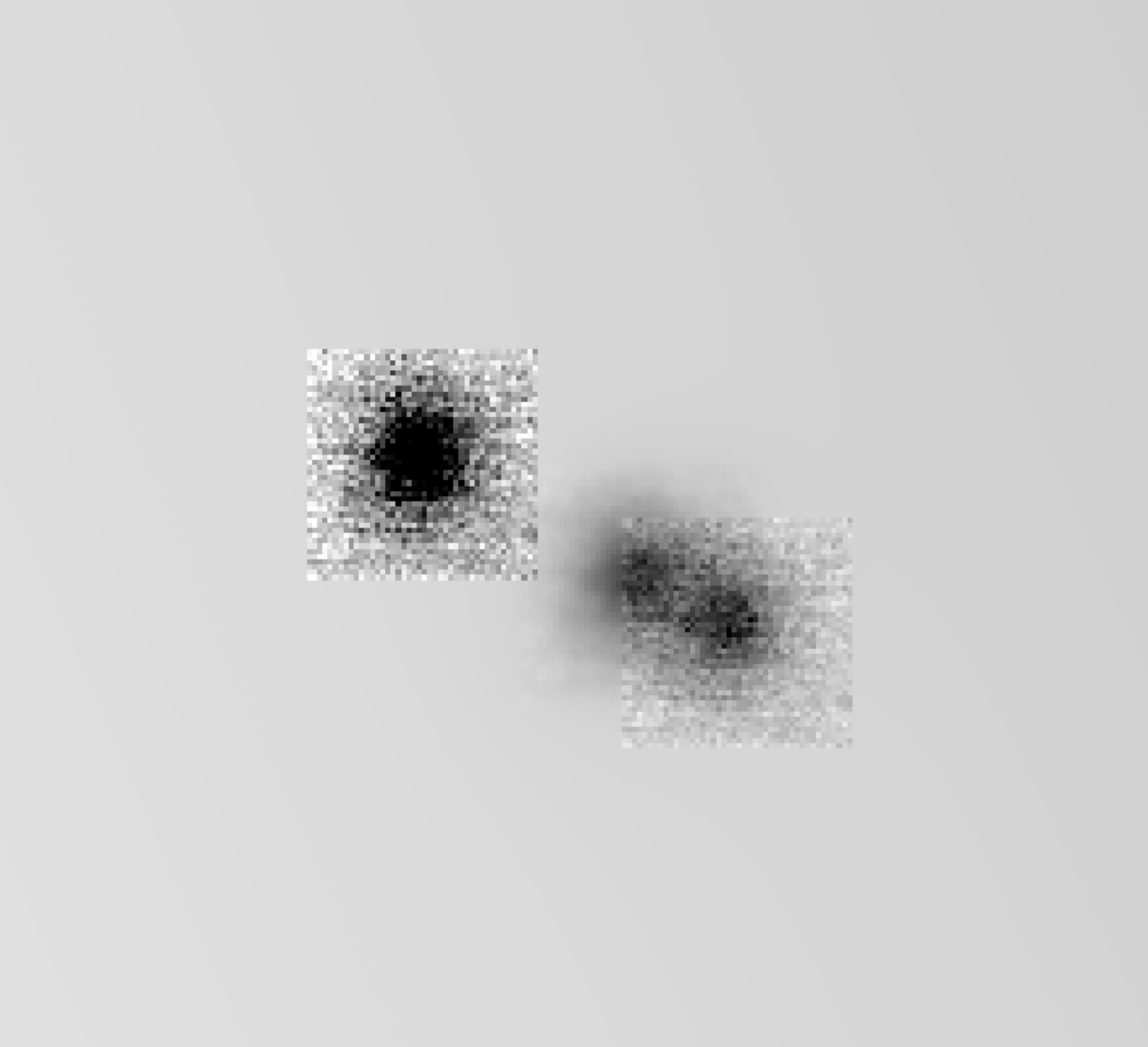}}
\fbox{\includegraphics[clip=true, trim=4cm 9cm 4.5cm 7cm, width=6.cm]{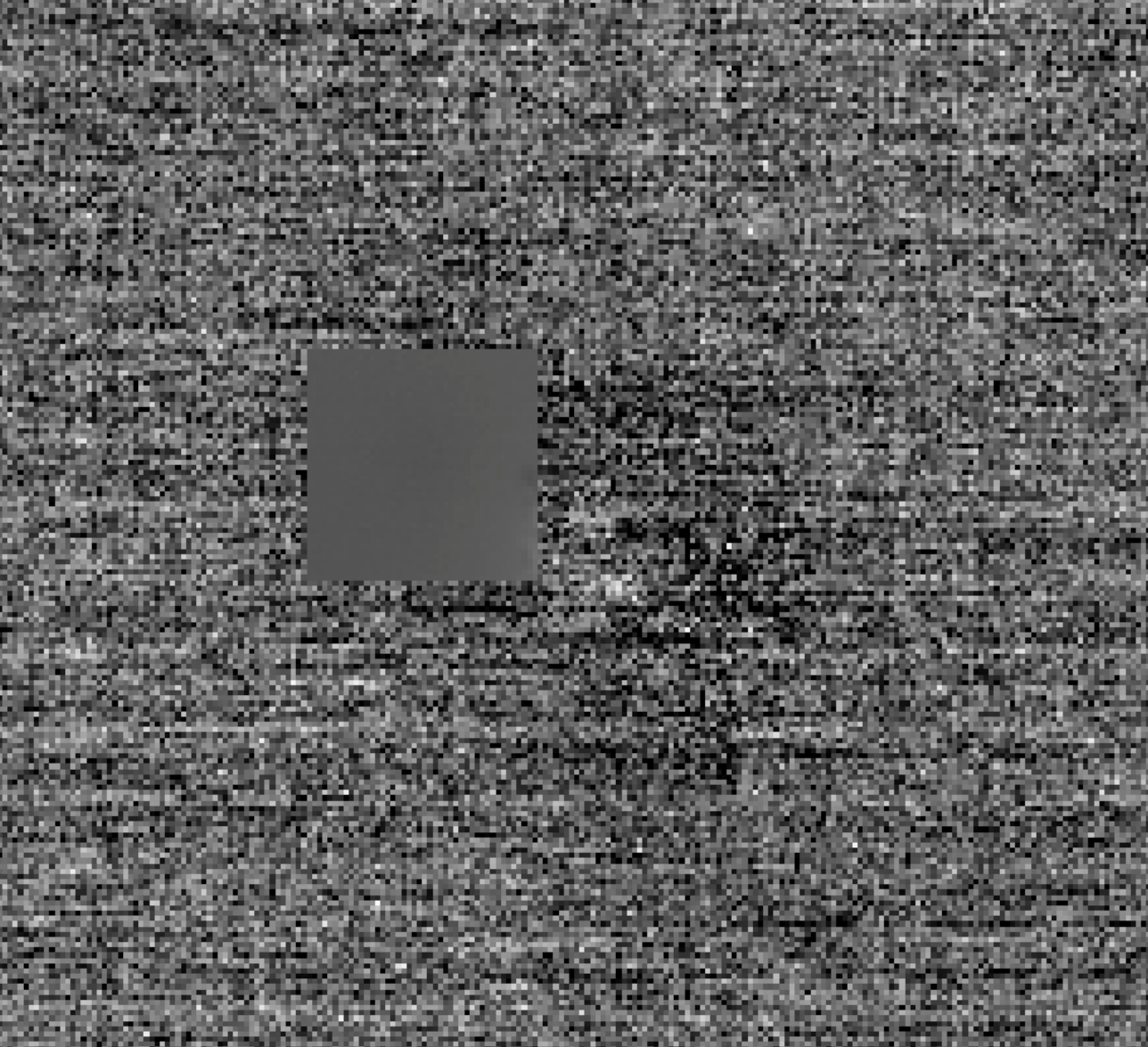}}
\caption{A photometric model for the lens SDSS 0806+2006 produced by our \texttt{GALFIT} analysis of the $K$-band image in Figure \ref{fig:data2}.  The top left panel shows the original data with the lens galaxy and quasar images labeled as G, A, and B, respectively.  Image A provides the model PSF.  The top right panel shows the model for the various components, while the bottom panel shows the residuals after model subtraction, which follow a Gaussian distribution and have reduced $\chi_r^2=1$.}
\label{fig:galfit}
\end{figure*}

We follow a procedure similar to that of \citet{2008A&A...492L..39S}.  We select the lensed quasar image farthest from the lens galaxy as our initial model for the point spread function (PSF).\footnote{If the quasar images lie at similar distances from the lens galaxy, we select the brightest image.}  We then perform a simultaneous fit to all the lensed quasar images (modeled as scaled PSFs) and any galaxies present (modeled as PSF-convolved S\'ersic profiles).  With this fit in hand, we subtract the galaxy (or galaxies) from the original image and then re-exact a new PSF model that contains less contamination from the lens galaxy.  Using this cleaner PSF, we perform a new simultaneous fit.  We compute the reduced $\chi_r^2$ value for the residuals using regions defined by the PSF size for the quasar images, and regions that are 1, 2, or 3 times the PSF size for the lens galaxy.  If $\chi_r^2=1$ within these regions, we conclude that the procedure has converged.  If not, we perform additional iterations, improving the model at each step, until convergence 
is achieved.  Our $K$-band images typically require 3 or 4 iterations, while our $L'$ images require only 1 or 2 since the lens galaxy is rarely detected in $L'$.  \texttt{GALFIT} provides photometric errorbars, although each quasar image that serves as the PSF model is fit perfectly.  For these images, we determine the uncertainty by computing the range in flux where $\Delta \chi^2 =1$ in the PSF region.  (When the lensed quasar images are clearly distinct, we verify the derived flux values and uncertainties using aperture photometry.)  Figure \ref{fig:galfit} shows an example of the \texttt{GALFIT} modeling procedure, and Table \ref{tab:photometry} gives the derived flux ratios.

We calibrate the flux measurements using observations of standard stars taken immediately before or after science observations.  We consider both aperture photometry and \texttt{GALFIT} modeling with the star itself providing the PSF model.  Table \ref{tab:calibphoto} gives the calibrated photometry, along with a list of the standard stars we use.  We report magnitudes in the Vega system.

\begin{deluxetable}{cccc}
\tablewidth{0pt}
\tablecaption{$K$ and $L'$ Flux Ratios}
\tablehead{
Target & Image &$K$ & $L'$  }
\startdata
Q0142$-$100        & A & $\equiv 1.0$ & $\equiv 1.0$  \\
                                 & B & $0.128\pm0.002$ & $0.132\pm0.006$ \\
\hline
SDSS 0246$-$0825 & A & $\equiv 1.0$ & $\equiv 1.0$  \\
                                 & B & $0.258\pm0.015$ & $0.331\pm0.016$ \\
\hline
HE 0435$-$1223  & A  & $1.837\pm0.130$ & $1.706\pm0.129$ \\
                                 & B & $1.271\pm0.073$ & $0.991\pm0.074$ \\ 
                                 & C & $\equiv 1.0$ & $\equiv 1.0$ \\
                                 & D & $0.851\pm0.049$ & $0.809\pm0.102$ \\                                  
\hline
SDSS 0806+2006 & A & $\equiv 1.0$ & $\equiv 1.0$  \\
                                 & B & $0.406\pm0.030$ & $<0.164\,(3\sigma)$ \\
\hline
SBS 0909+523     & A & $\equiv 1.0$ & $\equiv 1.0$  \\
                                 & B & $0.973\pm0.028$ & $0.973\pm0.033$ \\
\hline
HE 2149$-$2745  & A & $\equiv 1.0$ & $\equiv 1.0$  \\
                                 & B & $0.280\pm0.006$ & $0.240\pm0.009$ \\
\enddata
\label{tab:photometry}
%\tablecomments{}
\end{deluxetable}

\begin{deluxetable}{ccccc}
\tablewidth{0pt}
\tablecaption{$K$ and $L'$ Photometry}
\tablehead{
Target & Image &$K$ & $L'$ &Standard }
\startdata
Q0142$-$100        & A & $14.61\pm0.01\pm0.009$ & $13.57\pm0.01\pm0.020$ & 1,2 -- SA 92-342 ($K,L'$)\\
                                 & B & $16.84\pm0.02\pm0.009$ & $15.77\pm0.05\pm0.020$ & \\
\hline
SDSS 0246$-$0825 & A & $15.36\pm0.01\pm0.008$ & $13.71\pm0.01\pm0.008$ & 1 -- GSPC S618-D ($K$)\\
                                 & B & $16.83\pm0.06\pm0.008$ & $14.91\pm0.05\pm0.008$ & 2 -- HD 22686 ($L'$)  \\
\hline
HE 0435$-$1223  & A  & $16.70\pm0.01\pm0.010$ & $15.00\pm0.02\pm0.010$ & 1,2 -- HD 289907 ($K,L'$)\\
                                 & B & $17.10\pm0.02\pm0.010$ & $15.59\pm0.05\pm0.010$ & \\ 
                                 & C & $17.36\pm0.05\pm0.010$ & $15.58\pm0.05\pm0.010$ &  \\
                                 & D & $17.68\pm0.05\pm0.010$ & $15.81\pm0.11\pm0.010$ & \\                                  
\hline
SDSS 0806+2006 & A & $16.95\pm0.01\pm0.021$ & $15.25\pm0.01\pm0.020$ & 1,2 -- GSPC P486-R ($K,L'$) \\
                                 & B & $17.93\pm0.08\pm0.021$ & $>17.21\,(3\sigma)$ & \\
\hline
SBS 0909+523     & A & $14.59\pm0.01\pm0.011$ & $13.30\pm0.01\pm0.013$ & 3 -- HST P091-D ($K$) \\
                                 & B & $14.57\pm0.03\pm0.011$ & $13.32\pm0.03\pm0.013$ & 2 -- HD 84800 ($L'$) \\
\hline
HE 2149$-$2745  & A & $14.08\pm0.01\pm0.016$ & $14.23\pm0.01\pm0.006$ & 1 -- EG 141 ($K$)\\
                                 & B & $15.46\pm0.02\pm0.016$ & $15.78\pm0.04\pm0.006$ & 2 -- GL 811.1 ($L'$)\\
\enddata
\label{tab:calibphoto}
\tablecomments{Vega magnitudes of lensed images.  The quoted uncertainties include contributions from the lensed images themselves (first) and the photometric standard (second).  Col.\ 5 lists the standard star used for each passband.  References:
1-- \citet{2006MNRAS.373..781L}, 
2-- \citet{2003MNRAS.345..144L},
3-- \citet{1998AJ....116.2475P}.
}
\end{deluxetable}

\begin{deluxetable}{cccccc}
\tablewidth{0pt}
\tablecaption{Previous Flux Ratio Measurements}
\tablehead{
Target & Filter & \multicolumn{3}{c}{$B/A$} & Ref. }
\startdata
Q0142$-$100            & HST F555W &  \multicolumn{3}{c}{$0.127\pm0.009$}  & 1  \\
				& HST F675W &  \multicolumn{3}{c}{$0.121\pm0.004$}  & 1  \\
				& Johnson $R$\tablenotemark\dag &  \multicolumn{3}{c}{$0.146\pm0.009$}  & 2  \\
				& HST F814W &  \multicolumn{3}{c}{$0.146\pm0.016$}  & 3  \\
				& HST F160W &  \multicolumn{3}{c}{$0.121\pm0.004$}  & 1  \\
\hline
SDSS 0246$-$0825 & Sloan $u$ &  \multicolumn{3}{c}{$0.319\pm0.033$}  & 4  \\
				& Sloan $g$ &  \multicolumn{3}{c}{$0.310\pm0.010$}  & 4  \\
				& HST F555W &  \multicolumn{3}{c}{$0.483\pm0.010$}  & 4  \\
				& Sloan $r$ &  \multicolumn{3}{c}{$0.328\pm0.011$}  & 4  \\
				& Sloan $i$ &  \multicolumn{3}{c}{$0.340\pm0.016$}  & 4  \\
				& HST F814W &  \multicolumn{3}{c}{$0.247\pm0.017$}  & 4  \\
				& HST F160W &  \multicolumn{3}{c}{$0.291\pm0.010$}  & 4  \\
				& Johnson $H$ &  \multicolumn{3}{c}{$0.302\pm0.062$}  & 4  \\
				& Johnson $K'$ &  \multicolumn{3}{c}{$0.290$}  & 4  \\
\hline
SDSS 0806+2006    & Johnson $V$ &  \multicolumn{3}{c}{$0.581\pm0.012$}  & 5  \\
				& Johnson $R$ &  \multicolumn{3}{c}{$0.673\pm0.014$}  & 5  \\
				& Johnson $I$ &  \multicolumn{3}{c}{$0.581\pm0.012$}  & 5  \\
				& 0.39--1.10 $\mu$m &  \multicolumn{3}{c}{$0.7$}  & 5  \\
				& 0.45--0.87 $\mu$m &  \multicolumn{3}{c}{$0.77$}  & 6  \\
				& \ion{C}{3} \& \ion{Mg}{2} BEL &  \multicolumn{3}{c}{$0.454$}  & 6  \\
				& Johnson $H$ &  \multicolumn{3}{c}{$0.649\pm0.013$}  & 5  \\
				& Johnson $H$ &  \multicolumn{3}{c}{$0.474\pm0.043$}  & 6  \\
				& Johnson $K'$ &  \multicolumn{3}{c}{$0.689$}  & 5  \\
\hline
SBS 0909+523     	& 0.3--8 keV &  \multicolumn{3}{c}{$3.125\pm0.030$}  & 7  \\
				& HST F555W &  \multicolumn{3}{c}{$0.444\pm0.077$}  & 1  \\
				& Sloan $r$\tablenotemark\dag &  \multicolumn{3}{c}{$0.551\pm0.012$}  & 8  \\
				& HST F814W &  \multicolumn{3}{c}{$0.705\pm0.073$}  & 1  \\
				& HST F160W &  \multicolumn{3}{c}{$0.887\pm0.029$}  & 1  \\
\hline
HE 2149$-$2745     	& Johnson $B$ &  \multicolumn{3}{c}{$0.236\pm0.009$}  & 9  \\
				& HST F555W &  \multicolumn{3}{c}{$0.209\pm0.015$}  & 1  \\
				& Johnson $V$ &  \multicolumn{3}{c}{$0.236\pm0.018$}  & 9  \\
				& Johnson $R$ &  \multicolumn{3}{c}{$0.231\pm0.006$}  & 9  \\
				& HST F814W &  \multicolumn{3}{c}{$0.238\pm0.005$}  & 1  \\
				& HST F160W &  \multicolumn{3}{c}{$0.238\pm0.009$}  & 1  \\
\hline
Target & Filter & $A/C$ & $B/C$ & $D/C$ & Ref. \\
\hline
HE 0435$-$1223      	& HST F555W &  $1.84\pm0.11$ & $1.08\pm0.09$ & $0.95\pm0.06$ & 10 \\  
				& KPNO $R$\tablenotemark\dag  & $1.75\pm0.10$ & $1.00\pm0.04$ & $0.85\pm0.05$ & 10 \\ 
				& Johnson $R$\tablenotemark\ddag  & $1.69\pm0.06$ & $0.95\pm0.01$ & $0.82\pm0.03$ & 11 \\ 
				& HST F814W &  $1.69\pm0.04$ & $1.01\pm0.04$ & $0.82\pm0.03$ & 10 \\ 
				& HST F160W &  $1.57\pm0.05$ & $1.00\pm0.03$ & $0.79\pm0.03$ & 10 \\  
\enddata
\tablenotetext\dag{Reported flux ratio is based on ground-based monitoring.  Uncertainty includes scatter due to intrinsic and/or microlensing variability.}
\tablenotetext\ddag{Reported flux ratio is based on ground-based monitoring.  Flux ratios are corrected for microlensing and intrinsic variabilty.}
\label{tab:existing}
\tablecomments{This is a non-comprehensive list of existing flux ratio measurements for our targets. References: 
1-- \citet{2000ApJ...536..584L},
2-- \citet{2010MNRAS.401.2805K},
3-- CASTLES (http://www.cfa.harvard.edu/castles/),
4-- \citet{2005AJ....130.1967I},
5-- \citet{2006AJ....131.1934I},
6-- \citet{2008A&A...492L..39S},
7-- \citet{2009ApJ...692..677D},
8-- \citet{2008NewA...13..182G},
9-- \citet{1998A&A...339L..13L},
10-- \citet{2006ApJ...640...47K},
11-- \citet{2010arXiv1009.1473C}.}
\end{deluxetable}

%%%%%%%%%%%%%%%%%%%%%%%%%%%%%%%%%%%%%%%%%%%%%%%%%%%%%%%%%%%%%%%%%

\section{Results}
\label{sec:irresults}

%%%%%%%%%%%%%%%%%%%%%%%%%%%%%%%%%%%%%%%%%%%%%%%%%%%%%%%%%%%%%%%%%

In this section we use our new measurements to look for differences between the $K$ and $L'$ flux ratios.  In the next section we compare the $K$ and $L'$ results with previous measurements at other wavelengths to discuss wavelength dependence more generally.

\begin{itemize}
\item \textit{Q0142$-$100.}
The $K$ and $L'$ flux ratios are $F_B/F_A=0.128\pm0.002$ and $0.132\pm0.006$, respectively, which are consistent within $1\sigma$.

\item \textit{SDSS 0246$-$0825.}
The flux ratios are $F_B/F_A=0.258\pm0.015$ and $0.331\pm0.016$ in the $K$ and $L'$ bands, respectively.  The difference is inconsistent with measurement noise at $>99.9\%$ confidence.

\item \textit{HE 0435$-$1223.}
We consider flux ratios with respect to image C.  For the A/C and D/C flux ratios, the $K$ and $L'$ values are consistent within the errorbars.  For the B/C flux ratio, however, there is a significant difference with $F_B/F_C=1.271\pm0.073$ in $K$ and $0.991\pm0.074$ in $L'$.  This discrepancy is significant at $>99\%$ confidence.

\item \textit{SDSS 0806+2006.}
In the $K$ band the flux ratio is $F_B/F_A=0.406\pm0.030$.  In the $L'$ band, however, we do not significantly detect image B, with an upper limit of $F_B/F_A<0.164$ ($3\sigma$).  This is the most striking wavelength dependence in our sample.

\item \textit{SBS 0909+523.}
We measure identical flux ratios of $F_B/F_A=0.973$ in $K$ and $L'$.  This null detection is interesting in comparison with known wavelength dependence in other bands (see \S \ref{sec:irdiscuss}).

\item \textit{HE 2149$-$2745.}
The flux ratios are $F_B/F_A=0.280\pm0.006$ and $F_B/F_A=0.240\pm0.009$ in $K$ and $L'$, respectively, which differ at $>99.99\%$ confidence.

\end{itemize}

%%%%%%%%%%%%%%%%%%%%%%%%%%%%%%%%%%%%%%%%%%%%%%%%%%%%%%%%%%%%%%%%%

\section{Discussion}
\label{sec:irdiscuss}

%%%%%%%%%%%%%%%%%%%%%%%%%%%%%%%%%%%%%%%%%%%%%%%%%%%%%%%%%%%%%%%%%

\begin{figure*}[!ht]
\centering
\includegraphics[clip=true, trim=3.5cm 1.cm 0.5cm 1.25cm, width=10.cm, angle=90]{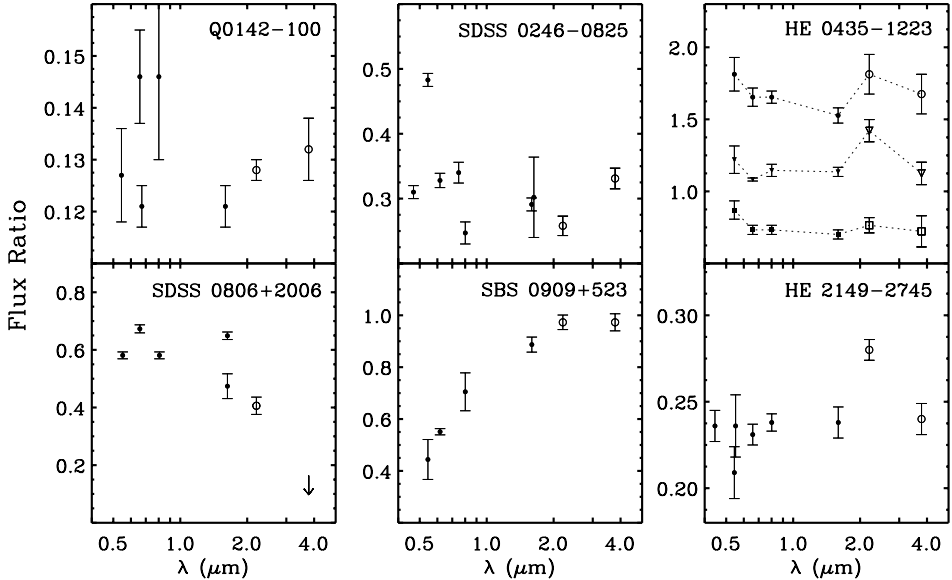}
\caption{Flux ratios for our lens sample across $0.4-3.8$ $\mu$m.  New NIR (Table \ref{tab:photometry}) and existing measurements (Table \ref{tab:existing}) are plotted with open and filled symbols, respectively.  For HE 0435-1223 we plot the flux ratios $F_A/F_C$, $F_B/F_C+0.2$, and $F_D/F_C$ from top to bottom, with a thin dotted line added for clarity.}
\label{fig:summary}
\end{figure*}

As discussed in \S \ref{sec:irintro}, wavelength dependence in lens flux ratios (especially between $L'$ and shorter wavelengths) may signal effects of dark matter substructure, but it may also be associated with quasar variability, microlensing, or dust extinction.  Here we compare our NIR flux ratio measurements (Table \ref{tab:photometry}) with previous measurements at other wavelengths (Table \ref{tab:existing}) in an attempt to disentangle the various effects that can alter flux ratios across wavelengths.  Even if we cannot uniquely identify the cause of wavelength dependence, our analysis highlights systems that have interesting features and warrant further study.  To ease comparison, we plot in Figure \ref{fig:summary} the flux ratios from Tables \ref{tab:photometry} and \ref{tab:existing} between $0.4-3.8$ $\mu$m.  

\textit{Q0142$-$100.}
Our $K$ and $L'$ flux ratios are compatible with each other and with single-epoch HST measurements in F555W, F675W, and F160W \citep{2000ApJ...536..584L}.  This is not surprising, because for a source redshift of $z_s=2.719$ all bands up to $L'$ probe rest-frame wavelengths $\le 1.0\,\mu$m where the accretion disk is the main emission source \citep[e.g.,][]{2008A&A...485...33H}.  What is striking is that $R$-band monitoring by \citet{2010MNRAS.401.2805K} yielded a mean flux ratio of $F_B/F_A=0.146$ and a scatter of 0.009 over 3-month observing windows in 2003--2005.  That value, which agrees with the single-epoch F814W measurement from CASTLES, suggests that there is more intrinsic/microlensing variability on several-year timescales than was apparent during the monitoring campaign.  

\textit{SDSS 0246$-$0825.}
Previous observations in various filters and epochs point to significant variations with both time and wavelength.  For instance, SDSS $i$ and HST F814W measurements taken 20 months apart yielded $F_B/F_A=0.340\pm0.016$ and $0.247\pm0.017$, respectively (\citealt{2005AJ....130.1967I}; CASTLES).  At the same two epochs, SDSS $g$ and HST  F555W measurements gave $F_B/F_A=0.310\pm0.010$ and $0.483\pm0.010$.  In the NIR, we find $F_B/F_A=0.258\pm0.015$ and $0.331\pm0.016$ in $K$ and $L'$.  It is unclear whether the NIR discrepancy is produced by the same variations that affect the shorter wavelengths, or whether millilensing by dark matter substructure might be at play.  To distinguish the possibilities, it would be interesting to use future multi-wavelength imaging to see whether optical and NIR flux ratios vary in similar ways.

\textit{HE 0435$-$1223.}
The measured A/C and D/C flux ratios are consistent with being independent of wavelength over 0.5--3.8 $\mu$m.  Our measurement of the B/C flux ratio in $L'$ is consistent with measurements at other wavelengths (including the $R$-band monitoring by \citealt{2006ApJ...640...47K}).  However, the $K$-band B/C flux ratio differs from the rest.  This is curious because $K$-band corresponds to rest-frame optical emission (0.82 $\mu$m at $z_s=1.689$), which should originate from the accretion disk.  Using the standard thin-disk model of \citet{1973A&A....24..337S}, we estimate the $K$ source size to be only 4.7 larger than the $R$ source size; for most plausible substructure models, such a difference is too small to account for the $K$-band anomaly \citep[see][]{2006MNRAS.365.1243D}.  Combining that with the agreement between the $R$ and $L'$ flux ratios, we conjecture that the $K$-band observations in 2008 may have been affected by microlensing.  Under this hypothesis, the $L'$-band observations were less sensitive to microlensing because the source was larger, while the earlier $R$-band monitoring happened to catch a period of smaller microlensing variations.  We study this interesting system in more detail in a separate paper (Fadely \& Keeton, in prep.).

\textit{SDSS 0806+2006.}
It is striking that we do not see image B in the $L'$-band observations ($F_B/F_A<0.164$ at $3\sigma$), especially considering that $F_B/F_A>0.4$ at shorter wavelengths.  Changes in the flux ratio with both wavelength and time (see Table \ref{tab:existing}) suggest that quasar variability and/or microlensing are at work in this lens (but not dust extinction; \citealt{2008A&A...492L..39S}).  It is not clear, though, whether either effect could explain the non-detection in $L'$.  For $z_s=1.54$ the $L'$ emission corresponds to rest-frame $1.5\,\mu$m, so it presumably contains some contribution from the dusty torus, and thus ought to be less susceptible to intrinsic variations or microlensing than the shorter-wavelength bands.  In any case, the wavelength dependence of the flux ratios is very interesting and should be investigated further with both modeling and follow-up observations.

\textit{SBS 0909+523.}
There is significant differential extinction in this lens \citep{2002ApJ...574..719M}; our observations should further constrain dust models since they represent the longest wavelengths yet measured in this system.  In fact, extrapolating the extinction models of \citeauthor{2002ApJ...574..719M} (which were derived from shorter-wavelength measurements) yields predicted flux ratios of $F_B/F_A=0.909$ in $K$ and 0.934 in $L'$, whereas we measure $0.97\pm0.03$ in both bands.  In terms of dark matter substructure, the agreement between our $K$ and $L'$ flux ratios is interesting as a null detection.  If the $K$ (source rest-frame $0.9\,\mu$m) and $L'$ (source rest-frame $1.6\,\mu$m) sources have different effective sizes, there must be no substructure near the images with the right Einstein radius to produce a chromatic effect.  Note that this does not rule out substructure in the lens --- only substructure with the right mass at the right position.
  
\textit{HE 2149$-$2745.}
The measured flux ratios are consistent with being independent of wavelength across 0.4--3.8 $\mu$m ($F_B/F_A \approx 0.24$), except that our $K$-band measurement ($F_B/F_A=0.280\pm0.006$) formally disagrees with the others.  However, the flux ratio is known to vary with recorded fluctuations at the level of 0.03 \citep{2002A&A...383...71B}.  Since $K$ band corresponds to rest-frame $0.7\,\mu$m, we conclude that the high $K$-band flux ratio is likely due to intrinsic and/or microlensing variability.

%%%%%%%%%%%%%%%%%%%%%%%%%%%%%%%%%%%%%%%%%%%%%%%%%%%%%%%%%%%%%%%%%

\section{Conclusions}

%%%%%%%%%%%%%%%%%%%%%%%%%%%%%%%%%%%%%%%%%%%%%%%%%%%%%%%%%%%%%%%%%

We have used a multi-wavelength analysis of lens flux ratios to search for effects of dark matter substructure, along with quasar variability, microlensing, and dust extinction.  We have presented new $K$ and $L'$ images of six lensed quasars, which were selected to have source redshifts $z_s<2.8$ so that $L'$ observations probe rest-frame wavelengths $>1.0\,\mu$m.  Some of the $L'$ flux should therefore originate from the extended torus of gas surrounding the central accretion disk, possibly providing the conditions for chromatic millilensing \citep[see][]{2006MNRAS.365.1243D}.

We find strong differences between the $K$ and $L'$ flux ratios for two lenses.  In HE 0435$-$1223 the $L'$ measurement of the B/C flux ratio is consistent with results at shorter wavelengths, but the $K$-band value is enhanced by $\sim$30\%.  We argue in a separate analysis (Fadely \& Keeton, in prep.) that the enhancement may be attributable to microlensing.  To test that hypothesis, future observations are needed to look for variability in the $K$-band and compare it with variability at shorter wavelengths.  In SDSS 0806+2006 we do not detect image B in $L'$ observations, with an upper limit of $F_B/F_A<0.164$ ($3\sigma$), even though the flux ratio is $F_B/F_A>0.4$ at shorter wavelengths.  It would be very interesting to use deeper observations at $L'$ to find image B, to measure the flux ratio at other long wavelengths, and to combine all the observations with lens modeling to ascertain whether the $L'$ anomaly may be caused by dark matter substructure or some other interesting effect.

We find a $\sim$30\% difference between the $K$ and $L'$ flux ratios for SDSS 0246$-$0825, and a smaller but statistically significant difference for HE 2149$-$2745.  Multi-epoch observations of both systems have shown variations in the flux ratios at shorter wavelengths, which are presumably associated with quasar variability and/or microlensing.  It is possible that the same phenomena affect the $K$-band flux ratios and explain the NIR discrepancies.  In that case, the flux ratios in $L'$ or other longer-wavelength bands would be useful for limiting effects of quasar variability and microlensing in order to probe dark matter substructure.  Alternatively, it remains possible that chromatic millilensing causes the NIR discrepancies.  To test both hypotheses, future observations are needed to quantify variability in multiple optical and NIR bands.

Finally, we detect no difference between the $K$ and $L'$ flux ratios for Q0142$-$100 and SBS 0909+523.  These results are interesting null detections, indicating that there is no substructure close to the images with masses in the right range to produce chromatic effects.  Such results offer constraints on dark matter substructure that may be mild but nevertheless interesting for future statistical studies of substructure populations.

\acknowledgements
We thank Gary Bernstein for helpful comments. This work was supported by NSF grant AST-0747311.

\end{document}